\documentclass{easychair}

\usepackage{hyphenat}
\usepackage{alltt}
\usepackage{fancyvrb}
\usepackage[figure,linesnumbered]{algorithm2e}

\newcommand{\assl}{ASSL\index{ASSL}\index{Autonomic Systems Specification Language}}
\newcommand{\marf}[0]{MARF\index{MARF}\index{Frameworks!MARF}\index{Libraries!MARF}}
\newcommand{\dmarf}[0]{DMARF\index{MARF!Distributed}\index{Frameworks!Distributed MARF}\index{Libraries!Distributed MARF}}
\newcommand{\admarf}[0]{ADMARF\index{ADMARF}\index{MARF!Autonomic}\index{DMARF!Autonomic}
            \index{Frameworks!Autonomic Distributed MARF}\index{Libraries!Autonomic Distributed MARF}}
\newcommand{\tool}[1]{\texttt{#1}\index{Tools!#1}}
\newcommand{\api}[1]{\url{#1}\index{API!#1}}
\newcommand{\xf}[1]{Figure~\ref{#1}}

\newcommand{\xs}[1]{Section~\ref{#1}}
\newcommand{\xa}[1]{Appendix~\ref{#1}}

\begin{document}

\title{
	Developing Autonomic Properties for Distributed Pattern-Recognition Systems with ASSL:
	A Distributed MARF Case Study
 \thanks{
  This work was supported in part by an IRCSET postdoctoral fellowship grant (now termed as EMPOWER) 
  at University College Dublin, Ireland, by the Science Foundation Ireland grant {03/CE2/I303\_1} to 
  Lero (the Irish Software Engineering Research Centre), and by the Faculty of Engineering and Computer 
  Science of Concordia University, Montreal, Canada.
 }
}

\titlerunning{Autonomic Specification of Pattern-Recognition Systems with ASSL}

\author{
 Emil Vassev\\
 \affiliation{Lero - the Irish Software Engineering Research Centre}\\
 \affiliation{University of Limerick, Limerick, Ireland}\\
 \url{emil@vassev.com}
 \and
 Serguei A. Mokhov\\ 
 \affiliation{Faculty of Engineering and Computer Science, Concordia University}\\
 \affiliation{1455 de Maisonneuve Blvd. W, Montreal, QC, Canada}\\
 \url{mokhov@cse.concordia.ca}
}

\authorrunning{Vassev and Mokhov}

\maketitle

\begin{abstract}
 In this paper, we discuss our research towards developing special properties that 
 introduce autonomic behavior in pattern-recognition systems. In our approach we use 
 ASSL (Autonomic System Specification Language) to formally develop such properties 
 for DMARF (Distributed Modular Audio Recognition Framework). These properties 
 enhance DMARF with an autonomic middleware that manages the four stages of the framework's 
 pattern-recognition pipeline. DMARF is a biologically inspired system employing pattern 
 recognition, signal processing, and natural language processing helping us process audio, 
 textual, or imagery data needed by a variety of scientific applications, e.g., biometric 
 applications. In that context, the notion go autonomic DMARF (ADMARF) can be employed by autonomous 
 and robotic systems that theoretically require less-to-none human intervention other than 
 data collection for pattern analysis and observing the results. In this article, we explain 
 the ASSL specification models for the autonomic properties of DMARF.\\
 {\bf Keywords:} autonomic computing, formal methods, ASSL, DMARF
\end{abstract}

\section{Introduction}
\label{sect:intro}

Today, we face the challenge of hardware and software complexity that appears to be the biggest
threat to the continuous progress in IT. Many initiatives towards complexity reduction in both
software and hardware have arisen with the advent of new theories and paradigms. Autonomic computing
(AC) \cite{autonomic-computing-2004} promises reduction of the workload needed to maintain complex
systems by transforming them into self-managing autonomic systems. The AC paradigm draws inspiration
from the human body's {\em autonomic nervous system} \cite{phorn-ac}. The idea is that software
systems can manage themselves and deal with dynamic requirements, as well as unanticipated threats,
automatically, just as the body does, by handling complexity through self-management.

Pattern recognition is a widely used biologically inspired technique in the modern computer science.
Algorithms for image and voice recognition have been derived from the human brain, which uses pattern
recognition to recognize shapes, images, voices, sounds, etc. In this research, we
applied the principles of AC to solve specific problems in distributed pattern-recognition systems, such as
availability, security, performance, etc. where the health of a distributed pipeline
is important. We tackled these issues by introducing self-management into the system behavior.
As a proof-of-concept (PoC) case study, we used {\assl} \cite{vassevASSLBook,vassevPhDThesis}
to develop the autonomic self-management properties for {\dmarf} \cite{dmarf-web-services-cisse08}, which is an
intrinsically complex pipelined distributed system composed of multi-level operational layers.
The ASSL framework
helped us develop the self-managing features first, which we then integrated into {\dmarf}. In
this paper, based on our results, we provide an attempt to generalize our experience on any similar-scale
distributed pipelined pattern-recognition system.

\subsection{Problem Statement and Proposed Solution}

Distributed {\marf} ({\dmarf}) could not be used in autonomous
systems of any kind as-is due to lack of provision for such a use
by applications that necessitate the self-management requirements.
Extending {\dmarf} directly to support the said requirements is a major
redesign and development effort to undertake for an open-source project.
Moreover, such and extended autonomic {\dmarf} must be validated and 
tested, since there is no immediate guarantee that the properties 
the latter has been augmented with are intrinsically correct.

In our approach, we provide the methodology for the initial proof-of-concept. 
We specify with {\assl} a number of autonomic properties for {\dmarf}, such as self-healing 
\cite{dmarf-assl-self-healing}, self-optimization \cite{dmarf-assl-self-optimization}, 
and self-protection \cite{dmarf-assl-self-protection}. The implementation of those 
properties is generated automatically by the {\assl} framework in the form of a 
special wrapper Java code that provides an autonomic layer implementing the 
{\dmarf}'s autonomic properties. In addition, the latter are formally validated 
with the {\assl}'s mechanisms for consistency and model checking. Model checking is 
performed on the generated Java code, where the ASSL relies on the Java PathFinder
\cite{mdl-chckng-java-pthfndr} tool developed by NASA Ames. 

\subsection{Organization}

The rest of this article is organized as follows. In \xs{sect:background}, we briefly review 
the field of autonomic computing and describe both {\assl} and {\dmarf} frameworks. 
\xs{sect:assl-dmarf-spec} presents details of the ASSL specification models for autonomic 
properties of {\dmarf}. Finally, \xs{sect:conclusion} presents some concluding remarks and 
future work.
 
\section{Background}
\label{sect:background}

The vision and metaphor of AC \cite{autonomic-computing-2004} is to apply the principles of 
self-regulation and complexity hiding. The AC paradigm emphasizes reduction of the workload 
needed to maintain complex systems by transforming those into self-managing autonomic systems 
(AS). The idea is that software systems shall automatically manage themselves just as the human 
body does. Nowadays, a great deal of research effort is devoted to developing AC tools. Such 
a tool is the {\assl} framework, which helps AC developers with problem specification, system 
design, system analysis and evaluation, and system implementation.

\label{page:review2-more-results} ASSL was initially developed by Vassev at
Concordia University, Montreal, Canada \cite{vassevPhDThesis} and since then
it has been successfully applied to the development of a variety of autonomic systems
including distributed ones. For example, ASSL was used to develop autonomic features
and generate prototype models for two NASA missions -- the ANTS (Autonomous Nano\hyp{}Technology Swarm)
concept mission (where a thousand of picospacecraft work cooperatively to explore the asteroid
belt \cite{truszkowski04}), and the Voyager mission \cite{voyager}. In both cases, there have been
developed autonomic prototypes to simulate the autonomic properties of the space exploration missions
and validate those properties through the simulated experimental results. The targeted autonomic
properties were: self-configuring \cite{assl-nasa-swarm-sac08}, self-healing \cite{assl-ants-slfhl},
and self-scheduling \cite{assl-ants-schdl} for ANTS, and autonomic image processing for Voyager \cite{assl-voyager}.
In general, the development of these properties required a two-level approach, i.e., they were specified
at the individual spacecraft level and at the level of the entire system. Because ANTS is intrinsically
distributed system composed of many autonomous spacecraft, that case study required individual
specification of the autonomic properties of each individual spacecraft member of the ANTS swarm.

\subsection{ASSL}
\label{sect:assl-intro}

The Autonomic System Specification Language ({\assl}) \cite{vassevASSLBook,vassevPhDThesis}
approaches the problem of formal specification and code generation of autonomic systems (ASs) 
within a framework. The core of this framework is a special formal notation and a toolset 
including tools that allow ASSL specifications be edited and validated. The current validation 
approach in {\assl} is a form of consistency checking (handles syntax and consistency errors) 
performed against a set of semantic definitions. The latter form a theory that aids in the 
construction of correct AS specifications. Moreover, from any valid specification, {\assl} 
can generate an operational Java application skeleton.

Overall, {\assl} considers autonomic systems (ASs) as composed of autonomic elements (AEs) 
communicating over interaction protocols. To specify those, {\assl} is defined through formalization 
of tiers. Over these tiers, {\assl} provides a multi-tier specification model that is designed to be 
scalable and exposes a judicious selection and configuration of infrastructure elements and mechanisms 
needed by an AS. The {\assl} tiers and their sub-tiers (see \xf{fig:assl-mt-model}) are abstractions 
of different aspects of the AS under consideration. They aid not only to specification of the system 
at different levels of abstraction, but also to reduction of the complexity, and thus, to improving 
the overall perception of the system. 

There are three major tiers (three major abstraction perspectives), each composed of sub-tiers 
(see \xf{fig:assl-mt-model}):

\begin{itemize}
\item
	{\em AS tier} -- presents a general and global AS perspective, where we define the general autonomic 
	system rules in terms of {\em service-level objectives (SLO)} and {\em self-management policies}, 
	{\em architecture topology} and {\em global actions}, {\em events} and {\em metrics} applied in these 
	rules. 
\item 
	{\em AS Interaction Protocol (ASIP) tier} -- forms a communication protocol perspective, where we 
	define the means of communication between AEs. An ASIP is composed of {\em channels},
	{\em communication functions}, and {\em messages}.
\item
	{\em AE tier} -- forms a unit-level perspective, where we define interacting sets of individual AEs 
	with their own behavior. This tier is composed of AE rules ({\em SLO} and {\em self-management policies}),
	an {\em AE interaction protocol} (AEIP), {\em AE friends} (a list of AEs forming a circle of trust), 
	{\em recovery protocols}, special {\em behavior models} and {\em outcomes}, {\em AE actions}, {\em AE events}, 
	and {\em AE metrics}.
\end{itemize}

\begin{figure}[t]
\begin{minipage}[t]{\columnwidth}
\begin{alltt}\scriptsize
\hrulefill
\textbf{I. Autonomic System (AS)}
  * \textbf{AS Service-level Objectives}
  * \textbf{AS Self-managing Policies}
  * \textbf{AS Architecture}
  * \textbf{AS Actions}
  * \textbf{AS Events}
  * \textbf{AS Metrics}
\textbf{II. AS Interaction Protocol (ASIP)}
  * \textbf{AS Messages}
  * \textbf{AS Communication Channels}
  * \textbf{AS Communication Functions}
\textbf{III. Autonomic Element (AE)}
  * \textbf{AE Service-level Objectives}
  * \textbf{AE Self-managing Policies}
  * \textbf{AE Friends}
  * \textbf{AE Interaction Protocol (AEIP)}
    - \textbf{AE Messages}
    - \textbf{AE Communication Channels}
    - \textbf{AE Communication Functions}
    - \textbf{AE Managed Elements}
  * \textbf{AE Recovery Protocol}
  * \textbf{AE Behavior Models}
  * \textbf{AE Outcomes}
  * \textbf{AE Actions}
  * \textbf{AE Events}
  * \textbf{AE Metrics} 
\hrulefill
\end{alltt}
\vspace{-2mm}
  \caption{ASSL Multi-Tier Model}
\vspace{-2mm}
  \label{fig:assl-mt-model}
\end{minipage}
\end{figure}

The AS Tier specifies an AS in terms of {\em service-level objectives} (AS SLOs), {\em self-management policies}, 
{\em architecture topology}, {\em actions}, {\em events}, and {\em metrics} (see \xf{fig:assl-mt-model}). 
The AS SLOs are a high-level form of behavioral specification that help developers establish system objectives 
(e.g., performance). The self-management policies could be any of (but not restricted to) the four so-called 
self-CHOP policies defined by the AC IBM blueprint: {\em self-configuring}, 
{\em self-healing}, {\em self-optimizing} and {\em self-protecting} \cite{ibmarchblprnt2006}.
These policies are event-driven
and trigger the execution of actions driving an AS in critical situations. The metrics constitute a 
set of parameters and observables controllable by an AS. At the ASIP Tier, the ASSL framework helps developers 
specify an AS-level interaction protocol as a public communication interface, expressed with special 
{\em communication channels}, {\em communication functions} and {\em communication messages}. At the AE Tier, 
the ASSL formal model exposes specification constructs for the specification of the system's AEs. 

\label{page:review2-more-results2} Conceptually, AEs are considered to be analogous to software agents
able to manage their own behavior and their relationships with other AEs. These relationships are specified
at both ASIP and AEIP tiers. Whereas ASIP specifies an AS-level {\em interaction protocol} that is public
and accessible to all the AEs of an AS and to {\em external systems} communicating with that very AS,
the AEIP tier is normally used to specify a {\em private communication protocol} used by an AE to communicate
only with: 1) trusted AEs, i.e., AEs declared as ``AE Friends'' (see \xf{fig:assl-mt-model}); and 2) special
controlled {\em managed elements}. Therefore, two AEs exchange messages over an AEIP only if they are
{\em friends}, thus revealing the need for special negotiation messages specified at ASIP to discover
new friends at runtime. 
 
Note that ASSL targets only the AC features of a system and helps developers clearly distinguish 
the AC features from the system-service features. This is possible, because with ASSL we model and generate special 
AC wrappers in the form of ASs that embed the components of non-AC systems. The latter are considered as 
{\em managed elements}, controlled by the AS in question. A managed element can be any software or hardware 
system (or sub-system) providing services. Managed elements are specified per AE (they form an extra layer at the 
AEIP see \xf{fig:assl-mt-model}) where the emphasis is on the control interface. It is important also to mention 
that the ASSL tiers and sub-tiers 
are intended to specify different aspects of an AS, but it is not necessary to employ all of them in order to 
model such a system. For a simple AS we need to specify 1) the AEs providing self-managing behavior intended to 
control the managed elements associated with an AE; and 2) the communication interface. Here, self-management 
policies must be specified to provide such self-managing behavior at the level of AS (the AS Tier) and at the 
level of AE (AE Tier). The self-management behavior of an ASSL-developed AS is specified with the self-management 
policies. These policies are specified with special ASSL constructs termed {\em fluents} and {\em mappings}
\cite{vassevASSLBook,vassevPhDThesis}. A fluent is a state where an AS enters with {\em fluent-activating events} 
and exits with {\em fluent\hyp{}terminating events}. A mapping connects fluents with particular actions to be 
undertaken. Usually, an ASSL specification is built around self-management policies, which make that specification
AC-driven. The policies themselves are driven by events and actions determined deterministically.
\xf{fig:self-management-plcy} presents a sample specification of an ASSL self-healing policy.

\begin{figure}[t] 
\begin{minipage}[t]{\columnwidth}
\begin{alltt} \scriptsize
\hrulefill
\textbf{ASSELF\_MANAGEMENT} \{ 
 \textbf{SELF\_HEALING} \{ 
  \textbf{FLUENT} inLosingSpacecraft \{ 
   \textbf{INITIATED_BY} \{ \textbf{EVENTS}.spaceCraftLost \}
   \textbf{TERMINATED_BY} \{ \textbf{EVENTS}.earthNotified \} 
	\} 
  \textbf{MAPPING} \{
   \textbf{CONDITIONS} \{ inLosingSpacecraft \}
   \textbf{DO_ACTIONS} \{ \textbf{ACTIONS}.notifyEarth \} 
  \}
 \}
\} \em{// ASSELF\_MANAGEMENT}
\hrulefill
\end{alltt}
\vspace{-2mm}
  \caption{Self-management Policy}
\vspace{-2mm}
  \label{fig:self-management-plcy}
\end{minipage}
\end{figure}

For more details on the {\assl} multi-tier specification model and the {\assl} framework toolset, please 
refer to \cite{vassevASSLBook,vassevPhDThesis}.

\subsection{Distributed MARF}
\label{sect:distr-dmarf-intro}

{\dmarf} \cite{dmarf-web-services-cisse08} is based on the classical {\marf}
whose pipeline stages were made into distributed nodes. The Modular Audio Recognition 
Framework ({\marf}) \cite{marf-ai08} is an open-source research platform 
and a collection of pattern recognition, signal processing, and natural language 
processing (NLP) algorithms written in Java and arranged into a modular and 
extensible framework facilitating addition of new algorithms for use and experiments 
by scientists. {\marf} can run distributively over the network, run stand-alone, 
or may just act as a library in applications. {\marf} has a number of algorithms 
implemented for various pattern recognition and some signal processing tasks. 
The backbone of {\marf} consists of pipeline stages that communicate with each 
other to get the data they need in a chained manner. 

\begin{figure*}[ht!]
	\begin{centering}
	\includegraphics[width=0.65\textwidth]{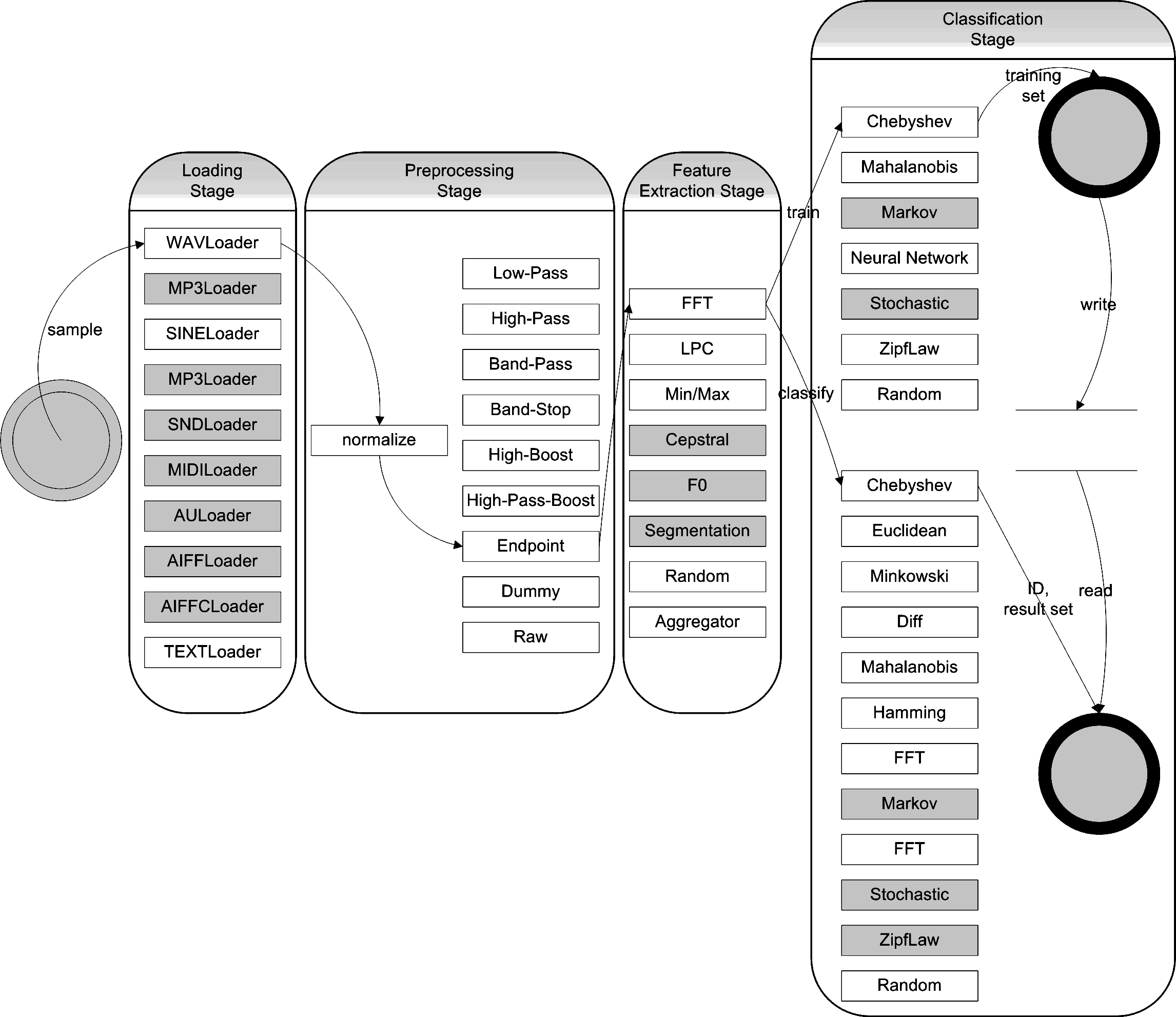}
	\caption{MARF's Pattern Recognition Pipeline}
	\label{fig:pipeline}
	\label{fig:pipeline-net}
	\end{centering}
\end{figure*}

In general, {\marf}'s pipeline of algorithm implementations is presented in \xf{fig:pipeline} 
(where the implemented algorithms are grouped in white boxes, and the stubs or in progress 
algorithms are grouped in gray). The pipeline consists of the four core stages grouping the 
similar kinds of algorithms: (1) sample loading, (2) preprocessing, (3) feature extraction, 
and (4) training/classification. {\marf}'s distributed extension, {\dmarf} \cite{dmarf-web-services-cisse08}
allows the stages of the pipeline to run as distributed nodes as well as a front-end.
The basic stages and the front-end were implemented without 
backup recovery or hot-swappable capabilities at this point; just communication over Java RMI 
\cite{java-rmi-short}, CORBA \cite{java-corba-idl-short}, and XML-RPC WebServices \cite{java-webservices-short}.
There is also an undergoing project on the intensional scripting language, MARFL\index{MARFL} 
\cite{marfl-context-secasa08} to allow scripting {\marf} tasks and applications.

There are various applications that test and employ {\marf}'s functionality and serve as examples of 
how to use {\marf}. High-volume processing of recorded audio, textual, or imagery data
are possible pattern-recognition and biometric applications of {\dmarf}. In this work, most of the emphasis 
is on audio processing, such as conference recordings with purpose of attribution of said 
material to identities of speakers. Another emphasis is on processing a bulk of recorded phone 
conversations in a police department for forensic analysis \cite{marf-file-type} and subject 
identification and classification. See the cited works and references therein for more
details on {\marf} and applications.

\section{Making Distributed Pipelined Systems Autonomic with ASSL: Autonomic DMARF Case Study}
\label{sect:assl-dmarf-spec}

\label{page:review1-generalization} In general, ASSL helps to design and generate special
{\em autonomic wrappers} in the form of AEs that embed one or more system components.
The latter are considered as {\em managed elements} (see \xs{sect:assl-intro}) 
that present one or more {\em single nodes} of a distributed system. Therefore, for each distributed node, we 
ideally specify with ASSL a single AE that introduces an autonomic behavior to that node. All the AEs are specified 
at the {\em AE Tier} and the global autonomic behavior of the entire system is handled by specifications at 
the {\em AS Tier} (see \xf{fig:assl-mt-model}). As shown in \xs{sect:assl-intro}, we rely on a rich set of constructs, 
such as {\em actions}, {\em events}, and {\em metrics} to specify special self-management policies driving the 
nodes of a distributed system in situations requiring autonomic behavior. Moreover, with ASSL, we specify 
special interaction protocols (ASIP and AEIP) that help those nodes exchange messages and synchronize on 
common autonomic behavior.
In this section we demonstrate how ASSL may be applied to an inherently distributed system such as {\dmarf}. The 
novelty in our approach is safeguarding the distributed pipeline, which is not possible with plain distributed 
systems. Therefore, with this case study we not only demonstrate the applicability of ASSL to distributed systems 
but also validate that ASSL may successfully be applied to {\em pipelined distributed systems}.

{\dmarf}'s capture as an AS primarily covers the autonomic behavior of the 
distributed pattern-recognition pipeline. We examine properties that apply to {\dmarf} and 
specify in detail the self-CHOP aspects of it. If we look a the {\dmarf} pipeline as a whole, 
we see that there should be at least one instance of every stage somewhere on the network. 
There are four main core pipeline stages and an application-specific stage that initiates pipeline 
processing. If one of the core stages goes offline, the pipeline stalls and to recover it has the 
following options: 1) use of a replacement node; 2) recovery of the failed node; or 3) rerouting 
the pipeline through a different node with the same service functionality as the failed one. 

In order to make {\dmarf} autonomic, we need to add {\em automicity} (autonomic computing behavior) 
to the {\dmarf} behavior. We add a special {\em autonomic manager} (AM) to each {\dmarf} stage. This 
makes the latter AEs, those composing an autonomic {\dmarf} ({\admarf}) capable of self-management.
 
\subsection{Self-Healing}

A {\dmarf}-based system should be able to recover itself through replication to keep at least one 
route of the pipeline available. There are two types of replication: 1) the replication of a service, 
which essentially means that we increase the number of nodes per core stage (e.g. two different hosts 
provide preprocessing services as active replication, so if one goes down, the pipeline is still 
not stalled; if both are up they can contribute to load balancing, which is a part of the self-optimization
autonomic property); and 2) replication within the node itself. If all nodes of a core stages go down, 
the stage preceding it is responsible to start up a temporary one on the host of the preceding stage, 
set it up to repair the pipeline. This is the hard replication needed to withstand stall faults, where 
it is more vulnerable and not fault-tolerant. In the second case, denoting passive replication of the 
same node (or even different nodes) losing a primary or a replica is not as serious as in the first 
case because such a loss does not produce a pipeline stall and it is easier to self-heal after a passive 
replica loss. Restart and recovery of the failed node without replicas is another possibility for 
self-healing for {\dmarf}. Technically, it may be tried prior or after the replica kicks in.

In the course of this project, we used {\assl} to specify the self-healing behavior of {\admarf} by 
addressing specific cases related to {\em node replacement} (service replica) and {\em node recovery},
shown in \xf{algo:self-healing}.

\begin{algorithm}[hptb]
\SetAlgoLined
{\admarf} monitors its run-time performance and in case of performance degradation notifies the problematic 
DMARF stages to start self-healing\;

Every notified DMARF stage (note that this is an AE) analyzes the problem locally to determine its nature: 
a node is down or a node is not healthy (does not perform well)\;

\If {A node is down} {
  \tcp{The following node-replacement algorithm is followed by the AM of the stage:}
  AM strives to find a replica note of the failed one\;
  \If {replica found} {next redirect computation to it\;}
  \If {replica not found} {report the problem. Note that the algorithm could be extended with a few more steps 
  where the AM contacts the AM of the previous stage to organize pipeline reparation\;}
}

\If {A node does not perform well} {
  \tcp{The following node-recovery algorithm is followed:}
  AM starts the recovery protocol for the problematic node\;
  \If {recovery successful} {do nothing\;}
  \If {recovery unsuccessful} {AM strives to find a replica node of the failed one\;}
  \If {replica found} {next redirect computation to it\;}
  \If {replica not found} {report the problem\;}
}
\caption{DMARF Self-Healing Algorithm}
\label{algo:self-healing}
\end{algorithm}

The following sub-sections describe the ASSL specification of the self-healing algorithm revealed here. 
We specified this algorithm as an ASSL self-healing policy spread on both system (AS tier) and autonomic 
element (AE tier) levels where {\em events}, {\em actions}, {\em metrics}, and special {\em managed element 
interface functions} are used to incorporate the self-healing behavior in ADMARF (see \xa{appdx:assl-self-healing-spec}).
Note that due to space limitations \xa{appdx:assl-self-healing-spec} presents a partial ASSL specification 
where only one AE (DMARF stage) is specified. The full specification specifies all the four DMARF stages.

\subsubsection{AS Tier Specification for Self-Healing}
\label{sect:as-tier-spec-heal}

At the AS tier we specify the global ADMARF self-healing behavior. To specify the latter, we use an ASSL 
\api{SELF_HEALING} self-management policy (see Figure \ref{fig:as-tier-self-healing}). Here we 
specified a single {\em fluent} mapped to an {\em action} via a {\em mapping}. 

\begin{figure}[t] 
\begin{minipage}[t]{\columnwidth}
\begin{alltt} \scriptsize
\hrulefill
\textbf{ASSELF\_MANAGEMENT} \{
  \textbf{SELF\_HEALING} \{
    {\em // a performance problem has been detected} 
    \textbf{FLUENT} inLowPerformance \{
      \textbf{INITIATED\_BY} \{ \textbf{EVENTS}.lowPerformanceDetected \}
      \textbf{TERMINATED\_BY} \{	\textbf{EVENTS}.performanceNormalized, 
                      \textbf{EVENTS}.performanceNormFailed \} 
    \}
    \textbf{MAPPING} \{
      \textbf{CONDITIONS} \{ inLowPerformance \} 
      \textbf{DO\_ACTIONS} \{ \textbf{ACTIONS}.startSelfHealing \}  
    \}
  \}  
\} {\em // ASSELF\_MANAGEMENT}
\hrulefill
\end{alltt}
\vspace{-2mm}
  \caption{AS Tier SELF\_HEALING Policy}
\vspace{-2mm}
  \label{fig:as-tier-self-healing}
\end{minipage}
\end{figure}

Thus, the \api{inLowPerformance} fluent is initiated by a \api{lowPerformanceDetected} event and 
terminated by one of the events such as \api{performanceNormalized} or \api{performanceNormFailed}.
Here the \api{inLowPerformance} event is activated when special AS-level \api{performance}
service-level objectives (SLO) degrade (see \xf{fig:as-tier-slo-events}). 
Note that in {\assl}, SLO are evaluated as {\em Booleans} based on their
{\em satisfaction} and thus, they can be evaluated as {\em degraded} or {\em normal} 
\cite{vassevPhDThesis}. Therefore, in our specification model, the \api{lowPerformanceDetected} 
event is activated anytime when the ADMARF's performance goes down. Alternatively, the 
\api{performanceNormalized} event activates when the same performance goes up.
 
\begin{figure}[t] 
\begin{minipage}[t]{\columnwidth}
\begin{alltt} \scriptsize
\hrulefill
\textbf{ASSLO} \{		
  \textbf{SLO} performance \{
    \textbf{FOREACH} member \textbf{in AES} \{
      member.\textbf{AESLO}.performance					
    \}				
  \}
\}
....
\textbf{EVENTS} \{ {\em // these events are used in the fluents specification} 
  \textbf{EVENT} lowPerformanceDetected \{ 
    \textbf{ACTIVATION} \{ \textbf{DEGRADED} \{ \textbf{ASSLO}.performance \} \} \}
  \textbf{EVENT} performanceNormalized \{ 
    \textbf{ACTIVATION} \{ \textbf{NORMALIZED} \{ \textbf{ASSLO}.performance \} \} \} 
  \textbf{EVENT} performanceNormFailed \{ 
    \textbf{ACTIVATION} \{ 
      \textbf{OCCURRED} \{ \textbf{AES.STAGE_AE.EVENTS}.selfHealingFailed \} \} \} 
\} {\em // EVENTS}
\hrulefill
\end{alltt}
\vspace{-2mm}
  \caption{AS Tier SLO and Events}
\vspace{-2mm}
  \label{fig:as-tier-slo-events}
\end{minipage}
\end{figure}

As specified, the AS-level \api{performance} SLO are a global task
whose realization is distributed among the AEs (DMARF stages).
Thus, the AS-level performance degrades when the performance of any
of the DMARF stages goes down (see the \api{FOREACH} loop in
\xf{fig:as-tier-slo-events}), thus triggering the \api{SELF_HEALING} policy.
In addition, the \api{performanceNormFailed} event activates if an a
special event (\api{selfHealingFailed}) occurs in the system.
This event is specified at the AE tier (see \xs{sect:ae-tier-spec-heal})
and reports that the local AE-level self-healing has failed.
Although not presented in this specification, the \api{performanceNormFailed}
event should be activated by any of the \api{performanceNormFailed}
events specified for each AE (a DMARF stage).

Moreover, once the \api{inLowPerformance} fluent gets initiated, the
corresponding \api{startSelfHealing} action is executed (see \xf{fig:as-tier-slo-events}).
This action simply triggers an AE-level event if the performance of that AE is degraded.
The AE-level event prompts the \api{SELF_HEALING} policy at the AE level (see \xs{sect:ae-tier-spec-heal}).

\subsubsection{AE Tier Specification for Self-Healing}
\label{sect:ae-tier-spec-heal}

At this tier we specify the self-healing policy for each AE in the ADMARF AS. Recall that the {\admarf}'s 
AEs are the DMARF  stages enriched with a special autonomic manager each. 

\xa{appdx:assl-self-healing-spec} presents the self-healing specification of one AE called \api{STAGE_AE}. 
Note that the latter can be considered as a generic AE and the specifications of the four AEs (one per DMARF 
stage) can be derived from this one.
Similar to the AS-level specification (see \xs{sect:as-tier-spec-heal}), here we specify a (but AE-level)
\api{SELF_HEALING} policy with a set of {\em fluents} initiated and terminated by {\em events} and 
{\em actions} mapped to those {\em fluents} (see \xa{appdx:assl-self-healing-spec}). Thus we specified 
three distinct fluents: \api{inActiveSelfHealing}, \api{inFailedNodesDetec}-\api{ted}, and 
\api{inProblematicNodesDetected}, each mapped to an AE-level action. The first fluent gets initiated 
when a \api{mustDoSelfHealing} event occurs in the system. That event is triggered by the AS-level 
\api{startSelfHea}-\api{ling} action in the case when the \api{performance} SLO of the AE get degraded 
(see \xa{appdx:assl-self-healing-spec}).  

Here the \api{performance} SLO of the AE are specified as a Boolean expression over two ASSL metrics, 
such as the \api{numberOfFailedNodes} metric and the equivalent \api{numberOfProblematicNodes} 
(see \xf{fig:ae-slo}).  Whereas the former measures the number of {\em failed notes} in the DMARF 
stage, the latter measures the number of {\em problematic nodes} in that stage.

\begin{figure}[t] 
\begin{minipage}[t]{\columnwidth}
\begin{alltt} \scriptsize
\hrulefill
\textbf{AESLO} \{		
  \textbf{SLO} performance \{
    \textbf{METRICS}.numberOfFailedNodes 
    \textbf{AND}
    \textbf{METRICS}.numberOfProblematicNodes
  \}
\}
\hrulefill
\end{alltt}
\vspace{-2mm}
  \caption{AE Tier SLO}
\vspace{-2mm}
  \label{fig:ae-slo}
\end{minipage}
\end{figure}

Both metrics are specified as \api{RESOURCE} metrics, i.e., observing a managed resource controlled 
by the AE \cite{vassevPhDThesis}. Note that the managed resource is the DMARF stage itself. Thus, 
as those metrics are specified (see \xa{appdx:assl-self-healing-spec}) they get updated by the DMARF 
stage via special {\em interface functions} embedded in the specification of a \api{STAGE_ME} managed 
element (see \xa{appdx:assl-self-healing-spec}). In addition, both metrics are set to accept only a 
zero value (see \xf{fig:ae-metric}), thus set in the so-called metric \api{THRESHOLD_CLASS} 
\cite{vassevPhDThesis}. The latter determines rules for {\em valid} and {\em invalid} metric values. 
Since in {\assl} metrics are evaluated as Booleans (valid or invalid) based on the value they are 
currently holding, the performance SLO (see \xf{fig:ae-slo}) gets degraded if one of the two defined 
metrics, the \api{numberOf}-\api{FailedNodes} metric or the \api{numberOfProblematicNodes} metric become 
{\em invalid}, i.e., if the DMARF stage reports that there is one or more {\em failed} or {\em problematic} nodes.

\begin{figure}[t] 
\begin{minipage}[t]{\columnwidth}
\begin{alltt} \scriptsize
\hrulefill
\textbf{VALUE} \{ 0 \}
\textbf{THRESHOLD\_CLASS} \{	\textbf{Integer} [0] \} {\em // valid only when holds 0}
\hrulefill
\end{alltt}
\vspace{-2mm}
  \caption{AE Tier Metric Threshold Class}
\vspace{-2mm}
  \label{fig:ae-metric}
\end{minipage}
\end{figure}

The \api{inActiveSelfHealing} fluent prompts the \api{ana}-\api{lyzeProblem} action execution (see
\xa{appdx:assl-self-healing-spec}). The latter uses the \api{STAGE_ME} managed element's interface 
functions to determine the nature of the problem -- is it a node that failed or it is a node that does 
not perform well. Based on this, the action triggers a \api{mustSwitch}-\api{ToNodeReplica} event 
or a \api{mustFixNode} event respectively. Each one of those events initiates a fluent in the AE 
\api{SELF_HEALING} policy to handle the performance problem. 
The \api{inFailedNodesDetected} fluent handles the case when a node has failed and its replica 
must be started and the \api{inProblematicNodesDetected} fluent handles the case when a node must be 
recovered. Here the first fluent prompts the execution of the \api{start}-\api{ReplicaNode} action and the 
second prompts the execution of the \api{fixProblematicNode} action. Internally, both actions call 
interface functions of the \api{STAGE_ME} managed element. Note that those functions trigger erroneous 
events if they do not succeed (see \xf{fig:ae-me-self-healing}). Those events terminate fluents of the AE
\api{SELF_HEALING} policy (see \xa{appdx:assl-self-healing-spec}).

\begin{figure}[t] 
\begin{minipage}[t]{\columnwidth}
\begin{alltt} \scriptsize
\hrulefill
{\em // runs the replica of a failed node}
\textbf{INTERFACE_FUNCTION} runNodeReplica \{										
  \textbf{PARAMETERS} \{ DMARFNode	node \}
  \textbf{ONERR\_TRIGGERS} \{ \textbf{EVENTS}.nodeReplicaFailed \}
\}
{\em // recovers a problematic node}
\textbf{INTERFACE_FUNCTION} recoverNode \{										
  \textbf{PARAMETERS} \{ DMARFNode	node \}	
  \textbf{ONERR\_TRIGGERS} \{ \textbf{EVENTS}.nodeCannotBeFixed \}
\}
\hrulefill
\end{alltt}
\vspace{-2mm}
  \caption{AE Tier STAGE\_ME Functions}
\vspace{-2mm}
  \label{fig:ae-me-self-healing}
\end{minipage}
\end{figure}

 It is important to mention that the \api{inFailedNodes}-\api{Detected} fluent gets initiated when the
 \api{mustSwitchTo}-\api{NodeReplica} event occurs in the system. The latter is triggered by the 
 \api{analyzeProblem} action. 

\begin{figure}[t] 
\begin{minipage}[t]{\columnwidth}
\begin{alltt} \scriptsize
\hrulefill
\textbf{EVENT} mustSwitchToNodeReplica \{
  \textbf{ACTIVATION} \{	
    \textbf{OCCURRED} \{	\textbf{EVENTS}.nodeCannotBeFixed \}
  \}
\}
\hrulefill
\end{alltt}
\vspace{-2mm}
  \caption{Event mustSwitchToNodeReplica}
\vspace{-2mm}
  \label{fig:ae-event}
\end{minipage}
\end{figure}

Moreover, the same event activates, according to its specification (see \xf{fig:ae-event}), if a 
\api{nodeCannotBeFixed} event occurs in the system, which is due to the inability of the \api{recoverNode} 
interface function to recover the problematic node (see \xf{fig:ae-me-self-healing}). Therefore, 
if a node cannot be recovered the \api{inFailedNodesDetected} fluent will be initiated in an attempt 
to start the replica of that node. Note that this conforms to the self-healing algorithm presented in
\xs{sect:assl-dmarf-spec}.

\subsection{ASSL Self-Protection Model for DMARF}
\label{sect:assl-dmarf-self-protection-spec}

For scientific and research computing on a local network in a controlled lab environment, runs of 
{\dmarf} do not need to be protected against malicious alteration or denial of service. However, as 
soon as the researchers across universities need to cooperate (or police departments to share the audio 
data recognition or computing), various needs about security and protection arise about data and 
computation results integrity, confidentiality, and the fact they come from a legitimate source.
Therefore, {\em self-protection} of a {\dmarf}-based system is less important in the localized
scientific environments, but is a lot more important in global environments ran over the Internet, 
potentially through links crossing country borders. This is even more true if the data being worked 
on by a {\dmarf} installation are of a sensitive nature such as recordings of the phone conversations 
of potential terrorist suspects. Thus, we point out the general requirements for this autonomic 
property of {\dmarf}:

\begin{itemize}
\item
 For the self-protection aspect, the {\dmarf}-based systems should adhere to the specification where 
 each node proves its identity to other nodes participating in the pipeline as well as passive replicas. 
 This will insure the data origin authentication (the data are coming from a legitimate source) and will
 protect against spoofing of the data with distorted voice recordings or incorrect processed data at the 
 later stages of the pipeline. Thus, we ensure the trustworthiness of the distributed data being processed
 \cite{marf-gipsy-distributed-ispdc08}. This can be achieved by proxy certificates issued to the nodes 
 during the deployment and management phase. Each node digitally signs the outgoing data, with the signature 
 the recipient node can verify through the certificate of the sender signed by trusted authority. This is 
 in a way similar to how DNSSec \cite{dnssec} operates for the DNS names and servers by attaching a public 
 key to the host and IP pair signed by the higher-level domain or authority. The similar trust mechanism 
 is also important when {\dmarf} is used for scientific research installation that say crosses the boundaries
 of several Universities' network perimeters over the Internet while performing scientific computation -- 
 the bottom line the data coming from the pipeline stages should be trustworthy, i.e. correct.
\item
 The same proxy certificates can also help with the data privacy along public channels, especially when 
 identities of real people are involved, such as speakers, that cross the Internet. The system should protect
 itself from any falsification attempt by detecting it, halting the corresponding computation, and logging 
 and reporting the incident in a trustworthy manner.
\item
 Run-time communication protocol selection is a self-protection option that ensures availability in case the 
 default communication mechanism becomes unavailable (e.g. the default port of \tool{rmiregistry} becomes blocked 
 by a firewall) the participating nodes switch to XML-RPC over HTTP. The protection of self against Distributed 
 Denial of Service (DDoS) is a difficult problem, which extends onto protecting not only self to the best 
 available means but also the self's peers by avoiding flooding them by the self's own output of a compromised 
 node. While the DDoS attacks are very difficult to mitigate if a node is under attack, each node can protect 
 self and others by limiting the amount of outgoing traffic it, itself, produces when a compromise is suspected
 or too much traffic flood is detected.
\end{itemize}

For self-protection, {\dmarf}-based systems should adhere to the specification 
where each node proves its identity to other nodes participating in the pipeline as 
well as passive replicas. This will insure the data origin authentication (the data 
are coming from a legitimate source) and will protect against spoofing of the data 
with distorted voice recordings or incorrect processed data at the later stages of 
the pipeline. Thus, we ensure the trustworthiness of the distributed data being processed
\cite{marf-gipsy-distributed-ispdc08}. This can be achieved by proxy certificates 
issued to the nodes during the deployment and management phase. Each node digitally 
signs the outgoing data, with the signature the recipient node can verify through the
certificate of the sender signed by trusted authority. This is in a way similar to how 
DNSSec \cite{dnssec} operates for the DNS names and servers by attaching a public key 
to the host and IP pair signed by the higher-level domain or authority. The same proxy 
certificates can also help with the data privacy along public channels, especially when 
identities of real people are involved, such as speakers, that cross the Internet. The 
system should protect itself from any falsification attempt by detecting it, halting
the corresponding computation, and logging and reporting the incident in a trustworthy manner.

To provide self-protecting capabilities, {\dmarf} has to incorporate special autonomic computing 
behavior. To achieve that, similar to our related work on the self-healing and self-optimization 
models for {\dmarf} \cite{dmarf-assl-self-optimization,dmarf-assl-self-healing}, we add a special 
autonomic manager (AM) to each {\dmarf} stage. This converts the latter into AEs that compose the
autonomic {\dmarf} ({\admarf}) capable of self-management. Self-protecting is one of the 
self-management properties that must be addressed by {\admarf}. Here we use {\assl} to specify 
the self-protecting behavior of {\admarf} where incoming messages must be secure in order to be 
able to process them. Thus, if a message (public or private) is about to be received in the AS, 
the following self-protection algorithm is followed by the AM of the stage (AE level) for private 
messages or by the global AM (AS level) for public messages:

\begin{itemize}
 \item
 A message hook mechanism detects when a message (public or private) is about to be received.
 \item
 AM strives to identify the sender of that message by checking the embedded digital signature:
 \begin{itemize}
  \item
  If the message does not carry a digital signature then it is considered insecure.
  \item
  If the message carries a digital signature then its digital signature is checked:
  \begin{itemize}
   \item
   If the digital signature is recognized then the message is considered secure.
   \item
   If the digital signature is not recognized then the message is considered insecure.
   \end{itemize}
  \end{itemize}
 \item
 If the message is secure no restrictions are imposed over the IO operations and the message can 
 be processed further.
 \item
 If the message is insecure the message is simply discarded by blocking any IO operations over 
 that message.
\end{itemize}

The following sections describe the {\dmarf} specification of the
self-protecting algorithm revealed here. We specified this algorithm
as an {\assl} self-protecting policy spread on both system (AS tier)
and autonomic element (AE tier) levels where {\em events}, {\em actions},
{\em metrics}, and special {\em managed element interface functions}
are used to incorporate the self-protecting behavior in {\admarf}
(see \xa{appdx:assl-self-protection-spec}). In addition,
two interaction protocols -- a public (ASIP tier) and a private (AEIP tier),
are specified to provide a secure communication system used
by both {\dmarf} nodes and external entities to communicate. Note that
due to space limitations \xa{appdx:assl-self-protection-spec} presents
a partial {\assl} specification where only one AE ({\dmarf} stage)
is specified. The full specification specifies all the four {\dmarf} stages.

\subsubsection{IP Tiers Specification}
\label{sect:asip-aeip-tier-spec}

Recall that {\assl} specifies AEs as entities communicating via special
{\em interaction protocols} (see \xs{sect:assl-intro}). Note that all
the communication activities (sending and receiving messages),
all the communication channels, and all the communication entities
({\assl} messages) must be specified in order to allow both internal
and external entities to communicate. Hence, no entity can either
send or receive a message that is not an {\assl}-specified message
or use alternative mechanism of communication. Thus, for the needs of 
the self-protecting mechanism, we specified two communication protocols -- 
at the ASIP tier and at the AEIP tier (this is nested in the AE specification 
structure) (see \xs{sect:assl-intro}). Please refer to \xa{appdx:assl-self-protection-spec} 
for a complete specification of both protocols.

At the ASIP tier, we specified a single public message (called \api{publicMessage}),
a single sequential bidirectional public communication channel (called \api{public}-\api{Link}),
and two public communication functions; specifically \api{receivePublicMessages} and 
\api{sendPublicMessa}-\api{ges}. They are specified to receive and send public messages 
over the public channel. Here any message sent or received must be an instance of the 
ASSL-specified \api{publicMessage}. The latter has an embedded \api{Proxy}-\api{Certificate} 
parameter specified to carry the digital signature of the message sender (see 
\xa{appdx:assl-self-protection-spec}). This parameter plays a key role in the self-protecting 
behavior of {\admarf}. Every message sender must complete this parameter with its {\em proxy 
certificate} before sending the message or the latter will be discarded by the system.

Moreover, the mentioned communication functions (\api{receivePublicMessages} and \api{sendPublicMessages})
are the only ones specified in the entire AS able to process instances of the \api{publicMessage} 
{\assl} message. Thus, to process such a message both functions are equipped with a conditional clause 
to check if the message is secure (see \xf{fig:rcv-public-msg}).

\begin{figure}[t] 
\begin{minipage}[t]{\columnwidth}
\begin{alltt} \scriptsize
\hrulefill
{\em //receive public messages if the message is secure}	
\textbf{FUNCTION} receivePublicMessages \{		
  \textbf{DOES} \{	
    \textbf{IF} ( \textbf{AS.METRICS}.thereIsInsecurePublicMessage ) \textbf{THEN}					
      \textbf{MESSAGES}.publicMessage << \textbf{CHANNELS}.publicLink
    \textbf{END}
  \}
\}
\hrulefill
\end{alltt}
\vspace{-2mm}
  \caption{ASSL Specification of receivePublicMessages}
\vspace{-2mm}
  \label{fig:rcv-public-msg}
\end{minipage}
\end{figure}

\xf{fig:rcv-public-msg} shows the \api{receivePublicMessages} communication function.
As is depicted, in order to send a public message the \api{thereIsInsecurePublicMessage}
metric (see \xa{appdx:assl-self-protection-spec}) must be {\em valid}. The latter is updated
by the self-protecting policy and is considered {\em invalid} (its operational evaluation returns
\api{FALSE} \cite{vassevPhDThesis}) if the public message to be received is {\em insecure}.

Note that the specification of the AEIP tier is identical to that of the ASIP tier (see
\xa{appdx:assl-self-protection-spec}), but deals with private messages \cite{vassevPhDThesis}, 
i.e., external entities cannot send or receive such messages.

\subsubsection{AS Tier Specification for Self-Protection}
\label{sect:as-tier-spec-protect}

To protect the AS from insecure public messages we specified a self-management policy that handles 
the verification of any incoming public message. Thus, at this tier we specify a \api{SELF_PROTECTING} 
policy (one of the four self-CHOP policies \cite{autonomic-computing-2004}) to ensure protection from 
{\em insecure} public messages. Here we specified a single fluent mapped to an action via a mapping 
clause (see \xf{fig:slf-prtctng-plcy}).

\begin{figure}[t] 
\begin{minipage}[t]{\columnwidth}
\begin{alltt} \scriptsize
\hrulefill
\textbf{SELF_PROTECTING} \{
  {\em // a new incoming message has been detected}  
  \textbf{FLUENT} inSecurityCheck \{
    \textbf{INITIATED_BY} \{ \textbf{EVENTS}.publicMessageIsComing \}
    \textbf{TERMINATED_BY} \{	\textbf{EVENTS}.publicMessageSecure, 
                   \textbf{EVENTS}.publicMessageInsecure \} 
  \}
  \textbf{MAPPING} \{
    \textbf{CONDITIONS} \{ inSecurityCheck \} 
    \textbf{DO_ACTIONS} \{ \textbf{ACTIONS}.checkPublicMessage \}  
  \}
\}  
\hrulefill
\end{alltt}
\vspace{-2mm}
  \caption{AS Tier SELF\_PROTECTING Policy}
\vspace{-2mm}
  \label{fig:slf-prtctng-plcy}
\end{minipage}
\end{figure}

The \api{inSecurityCheck} fluent is initiated by a \api{pub}-\api{licMessageIsComing}
event and is terminated by one of the events, such as \api{publicMessageSecure}
or \api{public}-\api{MessageInsecure}. Here the \api{inSecurityCheck} fluent is activated
when an instance of the ASIP-specified \api{publicMessage} is sent to a recipient in the AS.
Recall that any public message to be sent to a system recipient (e.g., a {\dmarf} node)
must be an instance of the {\assl} \api{publicMessage} message (see \xs{sect:asip-aeip-tier-spec}).
Therefore, in our specification model, the \api{publicMessageSecure} event will be activated
anytime when a \api{publicMessage} is about to be received by the AS (by a recipient in that AS).

\begin{figure}[t] 
\begin{minipage}[t]{\columnwidth}
\begin{alltt} \scriptsize
\hrulefill
\textbf{EVENTS} \{ {\em//these events are used in the fluents specification} 
  \textbf{EVENT} publicMessageIsComing \{ 
    \textbf{ACTIVATION} \{ \textbf{SENT} \{ \textbf{ASIP.MESSAGES}.publicMessage \} \} 
  \}
  \textbf{EVENT} publicMessageInsecure \{ 
    \textbf{GUARDS} \{ \textbf{NOT METRICS}.thereIsInsecurePublicMessage \}
    \textbf{ACTIVATION} \{ 
      \textbf{CHANGED} \{ \textbf{METRICS}.thereIsInsecurePublicMessage \} \}
  \} 
  \textbf{EVENT} publicMessageSecure \{ 		
    \textbf{GUARDS} \{ \textbf{METRICS}.thereIsInsecurePublicMessage \}
    \textbf{ACTIVATION} \{ 
      \textbf{CHANGED} \{ \textbf{METRICS}.thereIsInsecurePublicMessage \} \}
  \} 
\} {\em // EVENTS}
\hrulefill
\end{alltt}
\vspace{-2mm}
  \caption{AS Tier Events}
\vspace{-2mm}
  \label{fig:as-events}
\end{minipage}
\end{figure}

\xf{fig:as-events} presents the specification of all the three events used to initiate and 
terminate the \api{inSecurity}-\api{Check} fluent. As it is depicted, both \api{publicMessageIn}-
\api{secure} and \api{publicMessageSecure} are prompted when \api{thereIsInsecurePublicMessage}'s
value has changed. Special \api{GUARDS} are specified to prevent those events be prompted 
when that metric is {\em valid} or {\em not valid} respectively \cite{vassevPhDThesis}.
The corresponding metric \api{thereIsIn}-\api{securePublicMessage} accepts only Boolean values and is 
valid when it holds \api{FALSE}. The same metric is set to \api{TRUE} or \api{FALSE} by the 
\api{checkPublicMessage} action. Here the metric is set to \api{TRUE} anytime when a new public 
insecure message has been discovered (see \xa{appdx:assl-self-protection-spec}).

The \api{checkPublicMessage} action is mapped to the \api{inSecurityCheck} fluent (see 
\xf{fig:slf-prtctng-plcy}). Here this action is performed anytime when the AS enters in a 
{\em security check} state (determined by the \api{inSecurityCheck} fluent). This action 
is intended to check how secure is the incoming \api{publicMessage}, which triggers the self-protecting
policy by prompting the \api{publicMessageIs}-\api{Coming} event (see \xf{fig:as-events}).
To do that, the \api{checkPub}-\api{licMessage} action calls for each AE in the AS a
\api{check}-\api{SenderCertificate} action that must be specified in each AE ({\dmarf} stage) 
(see \xf{fig:as-action}).

\begin{figure}[t] 
\begin{minipage}[t]{\columnwidth}
\begin{alltt} \scriptsize
\hrulefill
senderIdentified = \textbf{false};
\textbf{FOREACH member in AES} \{	
  \textbf{IF} ( \textbf{NOT} senderIdentified ) \textbf{THEN}
    senderIdentified =
            \textbf{call member.ACTIONS}.checkSenderCertificate 
            (\textbf{ASIP.MESSAGES}.publicMessage.senderSignature)
  \textbf{END}
\};
\textbf{IF NOT} senderIdentified \textbf{THEN}
  {\em// makes the metric invalid and thus, triggers the attached}
  {\em// event and blocks all the operations with public messages}
  \textbf{set METRICS}.thereIsInsecurePublicMessage.\textbf{VALUE = true}			
\textbf{END}
\hrulefill
\end{alltt}
\vspace{-2mm}
  \caption{AS checkPublicMessage Action -- Partial Specification}
\vspace{-2mm}
  \label{fig:as-action}
\end{minipage}
\end{figure}

The \api{checkSenderCertificate} action returns \api{TRUE} if the \api{publicMessage} carries 
a valid digital signature (see \xa{appdx:assl-self-protection-spec}), i.e., the message is sent 
by a trusted sender ({\dmarf} node). As depicted by \xf{fig:as-action}, if one of the AE 
returns \api{TRUE}, then the \api{publicMessage} is considered secure; otherwise, it is 
considered insecure. If the message is insecure the \api{thereIsInsecurePublic}-\api{Message} 
metric is set to \api{TRUE}, which blocks the IO operations over this message 
(see \xs{sect:asip-aeip-tier-spec} and \xa{appdx:assl-self-protection-spec}).

\subsubsection{AE Tier Specification for Self-Protection}
\label{sect:ae-tier-spec-protect}

At this tier we specify the self-protecting mechanism for private messages.
Thus, for each AE ({\dmarf} stage) we specify a \api{SELF_PROTECTING} self-management policy
identical to the same policy specified at the AS tier (see \xs{sect:as-tier-spec-protect}).
Note that this policy deals with private messages specified at the AEIP tier
(the AE's private interaction protocol -- see \xa{appdx:assl-self-protection-spec}).

Therefore, similarly to the same policy specified for the AS tier, the AE-level \api{SELF_PROTECTING} 
policy is specified with a single \api{inSecurityCheck} fluent mapped to a \api{checkPrivateMessage} 
action. The \api{inSecurity}-\api{Check} fluent is initiated by the \api{privateMessageIsCo}-\api{mming} 
event and terminated by the \api{privateMessageIs}-\api{Secure} event or by the \api{privateMessageSecure} 
event. These events are similar to their homologous events specified at the AS tier 
(see \xs{sect:as-tier-spec-protect}), but dealing with the AEIP-specified \api{privateMessage} 
message and with the \api{thereIsInsecurePrivateMessage} metric at the AE-level 
(see \xa{appdx:assl-self-protection-spec}).

To perform the security checks of incoming private messages, the \api{checkPrivateMessage}
action invokes the \api{checkSenderCertificate} action (recall that the same action is
called by the \api{checkPublicMessage} action to check public messages -- see \xs{sect:as-tier-spec-protect}). 
Internally, the \api{checkSenderCertificate} action calls a {\em managed element interface function} 
specified at the AEIP protocol to check proxy certificates (see \xf{fig:ae-me-self-protection}).

\begin{figure}[t] 
\begin{minipage}[t]{\columnwidth}
\begin{alltt} \scriptsize
\hrulefill
\textbf{MANAGED_ELEMENTS} \{
  \textbf{MANAGED_ELEMENT STAGE_ME} \{	
    {\em// checks if a node certificate is valid}
    \textbf{INTERFACE_FUNCTION} checkNodeCertificate \{										
      \textbf{PARAMETERS} \{ ProxyCertificate	theCertificate \}	
      \textbf{RETURNS} \{ \textbf{Boolean} \}	
    \}
  \}	
\}
\hrulefill
\end{alltt}
\vspace{-2mm}
  \caption{AE STAGE\_ME Managed Element}
\vspace{-2mm}
  \label{fig:ae-me-self-protection}
\end{minipage}
\end{figure}

Recall (see \xs{sect:assl-intro}) that managed elements provide special interface functions to control 
the {\dmarf} system. Hence, as depicted by \xf{fig:ae-me-self-protection}, we expect the {\dmarf} stage 
to verify whether a specific proxy certificate is valid and to return \api{TRUE} or \api{FALSE}.
{\dmarf} does that through the Java Data Security Framework (JDSF) \cite{jdsf-integrity-cisse08,jdsf-authentication-cisse08}.

\subsection{ASSL Self-Optimization Model for DMARF}
\label{sect:assl-dmarf-self-optimization-spec}

The two major functional requirements applicable to large {\dmarf} installations related to 
self-optimization are outlined below:

\paragraph{Training set classification data replication.} 
A {\dmarf}-based system may do a lot of multimedia data processing and number crunching throughout 
the pipeline. The bulk of I/O-bound data processing falls on the sample loading stage and the 
classification stage. The preprocessing, feature extraction, and classification stages also do a 
lot of CPU-bound number crunching, matrix operations, and other potentially heavy computations. 
The stand-alone local {\marf} instance employs dynamic programming to cache intermediate results, 
usually as feature vectors, inverse co-variance matrices, and other array-like data. 
A lot of this data is absorbed by the classification stages. In the case of the {\dmarf}, such data 
may end up being stored on different hosts that run the classification service potentially causing 
re-computation of the already computed data on other classification host that did a similar evaluation 
already. Thus, the classification stage nodes need to communicate to exchange the data they have lazily 
acquired among all the classification members. Such data mirroring/replication would optimize a lot of 
computational effort on the end nodes.

\paragraph{Dynamic communication protocol selection.}
Additional aspect of self-optimization is automatic selection of the available most efficient communication 
protocol. E.g., if {\dmarf} initially uses WebServices XML-RPC and later discovers all of its nodes can 
also communicate using say Java RMI, they can switch to that as their default protocol in order to avoid 
marshaling and de-marshaling heavy SOAP XML messages that are always a subject of a big overhead even in 
the compressed form.

Here, the DMARF Classification stage is augmented with a self-optimizing autonomic policy. We used ASSL 
to specify this  policy and generate implementation for the same. \xa{appdx:assl-self-optimization-spec} 
presents a partial specification of the ASSL self-optimization model for ADMARF. As specified, the autonomic 
behavior is encoded in a special ASSL construct denoted as \api{SELF_OPTIMIZING} policy. The latter is 
specified at two levels - the global AS-tier level and the level of single AE (the AE-tier). The algorithm 
behind is described by the following elements:

\begin{itemize}
 \item
 Any time when ADMARF enters in the Classification stage, a self-optimization behavior takes place. 
 \item
 The Classification stage itself forces the stage nodes synchronize their latest cached results. Here 
 each node is asked to get the results of the other nodes.
 \item
 Before proceeding with the problem computation, each stage node strives to adapt to the most efficient 
 currently available communication protocol. 
\end{itemize}

The following sections describe the ASSL specification of the self-optimization algorithm revealed here.

\subsubsection{AS Tier Specification for Self-Optimization}
\label{sect:as-tier-spec-opti}

At this tier we specify a system-level \api{SELF_OPTIMIZING} policy and the actions and events 
supporting that policy. As was mentioned, ASSL supports policy specification with special constructs called {\em fluents} 
and {\em mappings} \cite{vassevPhDThesis}. Whereas the former are special states with conditional 
duration, the latter map actions to be executed when the system enters in such a state. 

Figure \ref{fig:as-self-optmztn} depicts the AS-tier specification of the \api{SELF_OPTIMIZING} policy. 
As we see the policy is triggered when the special fluent \api{inClassificationStage} is initiated. 
Here when {\admarf} enters the Classification stage in its pipeline, an AS-level 
\api{enteringClassi}-\api{ficationStage} event is prompted to initiate the corresponding 
\api{inClassificationStage} fluent. 

\begin{figure}[t] 
\begin{minipage}[t]{\columnwidth}
\begin{alltt} \scriptsize
\hrulefill
\textbf{SELF_OPTIMIZING} \{
  {\em// DMARF enters in the Classification Stage} 
  \textbf{FLUENT} inClassificationStage \{
    \textbf{INITIATED_BY} \{ \textbf{EVENTS}.enteringClassificationStage \}
    \textbf{TERMINATED_BY} \{	\textbf{EVENTS}.optimizationSucceeded, 
                     \textbf{EVENTS}.optimizationNotSucceeded \} 
  \}
  \textbf{MAPPING} \{
    \textbf{CONDITIONS} \{ inClassificationStage \} 
    \textbf{DO_ACTIONS} \{ \textbf{ACTIONS}.runGlobalOptimization \}  
  \}
\}  
\hrulefill
\end{alltt}
\vspace{-2mm}
  \caption{AS Tier SELF\_OPTIMIZING Policy}
\vspace{-2mm}
  \label{fig:as-self-optmztn}
\end{minipage}
\end{figure}

Further, this fluent is mapped to an AS-level \api{run}-\api{GlobalOptimization} action (see
\xa{appdx:assl-self-optimization-spec}). This action iterates over all the Classification stage 
nodes specified as distinct AEs (see \xs{sect:ae-tier-spec-opti}) and calls for each node a special AE-level
\api{synchronizeResults} action (see \xa{appdx:assl-self-optimization-spec}). In case of exception, the
\api{optimizati}-\api{onNotSucceeded} event is issued; else the \api{optimizati}-\api{onSucceeded} event 
is issued. Both events terminate the \api{inClassificationStage} fluent, and consecutively ADMARF exits the 
\api{SELF_OPTIMIZING} policy.

To distinguish the AEs from the other AEs in ADMARF, we specified the architecture topology of the system. 
For this we used  the \api{ASARCHITECTURE} ASSL construct \cite{vassevPhDThesis}. \xa{appdx:assl-self-optimization-spec}
presents the specification of the ADMARF architecture topology. Note that this is a partial specification 
depicting only two AEs. The full \api{ASARCHITECTURE} specification includes all the AEs of ADMARF. As depicted, 
we specified a special group of AEs called \api{CLASSF_STAGE} with members all the AEs representing the 
Classification stage nodes. This group allows the \api{runGlobalOptimization} action iterates over the stage nodes.

\subsubsection{AE Tier Specification for Self-Optimization}
\label{sect:ae-tier-spec-opti}

At this tier we specified the \api{SELF_OPTIMIZING} policy for the Classification stage nodes. Here we specified 
for every node a distinct AE. (see \xa{appdx:assl-self-optimization-spec}) presents the partial specification of 
two AEs, each representing a single node of the Specification stage. At this level, self-optimization concentrates 
on adapting the single nodes to the most efficient communication protocol. Similar to the AS-level policy specification
(see \xs{sect:as-tier-spec-opti}), an \api{inCPAdaptation} fluent is specified to trigger such adaptation when ADMARF
enters in the Specification stage. This fluent is initiated by the AS-level \api{enteringClassificationStage} event.

The same fluent is mapped to an \api{adaptCP} action to perform the needed adaptation. This action is specified as
\api{IMPL}, i.e., requiring further implementation \cite{vassevPhDThesis}. In ASSL, we specify \api{IMPL} actions to 
hide complexity via abstraction. Here, the \api{adaptCP} action is a complex structure, which explanation is beyond the
scope of this paper. Therefore, we abstracted the specification of this action (through \api{IMPL}) and provided only 
the prerequisite {\em guard} conditions and prompted events. 

\vspace{-5pt}
\section{Conclusion}
\label{sect:conclusion}
\vspace{-5pt}

In this article, we have presented ASSL specification models for autonomic features of ADMARF. To develop these 
features, we devised algorithms with ASSL for the pipelined stages of the DMARF's pattern recognition pipeline. The 
autonomic features were specified as special self-managing policies for self-healing, self-protecting, and 
self-optimizing in ADMARF. The ADMARF system (upon completion of the open-source implementation)
will be able to fully function in autonomous
environments, be those on the Internet, large multimedia processing farms, robotic spacecraft that do their own 
analysis, or simply even pattern-recognition research groups that can rely more on the availability of their 
systems that run for multiple days, unattended. Although not a fully complete specification model for ADMARF, we have 
attempted to provide didactic evidence of how ASSL can help us achieve desired automicity in {\dmarf}.

Future work is concerned with further ADMARF development by including new autonomic features. For example, together 
with the full implementation and testing of the presented specification models, we intend to develop autonomic 
features covering the self-configuration aspects of ADMARF. These will help to construct an intelligent {\admarf}
system able to react automatically to dynamic requirements by finding possible solutions and applying those with 
no human interaction.

\vspace{-5pt}
\bibliographystyle{abbrv}
\bibliography{assl-dmarf-soft-sys-arxiv}

\appendix
\section{ASSL Specification of DMARF Self-Healing}
\label{appdx:assl-self-healing-spec}

What follows is the complete initial {\assl} specification of the self-healing aspect for the {\dmarf}'s pipeline stages
\cite{dmarf-assl-self-healing}.

\scriptsize
\begin{Verbatim}
// ASSL self-healing specification model for DMARF 

AS DMARF {  

	TYPES {	DMARFNode }

	ASSLO {		
		SLO performance {
			FOREACH member in AES {
				member.AESLO.performance					
			}				
		}
	}

	ASSELF_MANAGEMENT {
		SELF_HEALING {
		  // a performance problem has been detected 
			FLUENT inLowPerformance {
				INITIATED_BY { EVENTS.lowPerformanceDetected }
				TERMINATED_BY {	EVENTS.performanceNormalized, EVENTS.performanceNormFailed } 
			}
			MAPPING {
				CONDITIONS { inLowPerformance } 
				DO_ACTIONS { ACTIONS.startSelfHealing }  
			}
		}  
	} // ASSELF_MANAGEMENT
	
	ACTIONS {
		ACTION IMPL startSelfHealing {	
			GUARDS { ASSELF_MANAGEMENT.SELF_HEALING.inLowPerformance }
			TRIGGERS { 
				IF NOT AES.STAGE_AE.AESLO.performance THEN
					 AES.STAGE_AE.EVENTS.mustDoSelfHealing 
				END
			}
		} 
   } // ACTIONS

	EVENTS { // these events are used in the fluents specification 
		EVENT lowPerformanceDetected { ACTIVATION { DEGRADED { ASSLO.performance} } }
		EVENT performanceNormalized { ACTIVATION { NORMALIZED { ASSLO.performance} } } 
		EVENT performanceNormFailed { ACTIVATION { OCCURRED { AES.STAGE_AE.EVENTS.selfHealingFailed } } } 
	} // EVENTS

} // AS DMARF

AES {

	AE STAGE_AE {	
	
		VARS { DMARFNode nodeToRecover }

		AESLO {		
			SLO performance {
				METRICS.numberOfFailedNodes 
				AND
				METRICS.numberOfProblematicNodes
			}
		}

		AESELF_MANAGEMENT {
			SELF_HEALING {
				FLUENT inActiveSelfHealing {
					INITIATED_BY { EVENTS.mustDoSelfHealing }
					TERMINATED_BY { EVENTS.selfHealingSuccessful, EVENTS.selfHealingFailed } 
				}
				FLUENT inFailedNodesDetected {
					INITIATED_BY { EVENTS.mustSwitchToNodeReplica }
					TERMINATED_BY { EVENTS.nodeReplicaStarted, EVENTS.nodeReplicaFailed } 
				}
				FLUENT inProblematicNodesDetected {
					INITIATED_BY { EVENTS.mustFixNode }
					TERMINATED_BY { EVENTS.nodeFixed, EVENTS.nodeCannotBeFixed } 
				}
				MAPPING {
					CONDITIONS { inActiveSelfHealing } 
					DO_ACTIONS { ACTIONS.analyzeProblem }  
				} 
				MAPPING {
					CONDITIONS { inFailedNodesDetected } 
					DO_ACTIONS { ACTIONS.startReplicaNode }  
				} 
				MAPPING {
					CONDITIONS { inProblematicNodesDetected } 
					DO_ACTIONS { ACTIONS.fixProblematicNode }  
				} 
			}  
		} // AESELF_MANAGEMENT

		AEIP {
			MANAGED_ELEMENTS {
				MANAGED_ELEMENT STAGE_ME {	
					INTERFACE_FUNCTION countFailedNodes {										
						RETURNS { Integer	}	
					}
					INTERFACE_FUNCTION countProblematicNodes {										
						RETURNS { Integer	}	
					}
					// returns the next failed node
					INTERFACE_FUNCTION getFailedNode {										
						RETURNS { DMARFNode	}	
					}
					// returns the next problematic node
					INTERFACE_FUNCTION getProblematicNode {										
						RETURNS { DMARFNode	}	
					}
					// runs the replica of a failed nodee
					INTERFACE_FUNCTION runNodeReplica {										
						PARAMETERS { DMARFNode	node }	
						ONERR_TRIGGERS { EVENTS.nodeReplicaFailed }
					}
					// recovers a problematic node
					INTERFACE_FUNCTION recoverNode {										
						PARAMETERS { DMARFNode	node }	
						ONERR_TRIGGERS { EVENTS.nodeCannotBeFixed }
					}
				}	
			}
		} // AEIP

		ACTIONS {
			ACTION analyzeProblem {						
				GUARDS { AESELF_MANAGEMENT.SELF_HEALING.inActiveSelfHealing }
				VARS { BOOLEAN failed }
				DOES { 
					IF METRICS.numberOfFailedNodes THEN
						AES.STAGE_AE.nodeToRecover = call AEIP.MANAGED_ELEMENTS.STAGE_ME.getFailedNode;
						failed = TRUE
					END
					ELSE
						AES.STAGE_AE.nodeToRecover = call AEIP.MANAGED_ELEMENTS.STAGE_ME.getProblematicNode;
						failed = FALSE
					END
				}
				TRIGGERS { 
					IF failed THEN EVENTS.mustSwitchToNodeReplica END
					ELSE EVENTS.mustFixNode END
				}
				ONERR_TRIGGERS { EVENTS.selfHealingFailed }
			} 

			ACTION startReplicaNode {						
				GUARDS { AESELF_MANAGEMENT.SELF_HEALING.inFailedNodesDetected }
				DOES {	call AEIP.MANAGED_ELEMENTS.STAGE_ME.runNodeReplica(AES.STAGE_AE.nodeToRecover) }
				TRIGGERS {	 EVENTS.nodeReplicaStarted }
			}

			ACTION fixProblematicNode {						
				GUARDS { AESELF_MANAGEMENT.SELF_HEALING.inProblematicNodesDetected }
				DOES {	call AEIP.MANAGED_ELEMENTS.STAGE_ME.recoverNode(AES.STAGE_AE.nodeToRecover) }
				TRIGGERS {	 EVENTS.nodeFixed}
			}
		} // ACTIONS

		EVENTS {
			EVENT mustDoSelfHealing { } 
			EVENT selfHealingSuccessful { 
				ACTIVATION {	
					OCCURRED {	 EVENTS.nodeReplicaStarted }  
					OR 
					OCCURRED {	 EVENTS.nodeFixed }
				} 
			}
			EVENT selfHealingFailed {	
				ACTIVATION {	
					OCCURRED {	 EVENTS.nodeReplicaFailed }  
				} 
			}
			EVENT mustSwitchToNodeReplica {
				ACTIVATION {	
					OCCURRED {	 EVENTS.nodeCannotBeFixed }
				}
			}
			EVENT nodeReplicaStarted { }
			EVENT nodeReplicaFailed {	}
			EVENT mustFixNode { }
			EVENT nodeFixed {	}
			EVENT nodeCannotBeFixed {	}
		} // EVENTS

		METRICS {				
			// increments when a failed node has been discovered
			METRIC numberOfFailedNodes {	
				METRIC_TYPE { RESOURCE	} 	
				METRIC_SOURCE {	 AEIP.MANAGED_ELEMENTS.STAGE_ME.countFailedNodes }
				DESCRIPTION {"counts failed nodes in the MARF stage"}	
				VALUE { 0 }
				THRESHOLD_CLASS {	Integer [0] } // valid only when holding 0 value
			}
			// increments when a problematic node has been discovered
			METRIC numberOfProblematicNodes {	
				METRIC_TYPE { RESOURCE	} 	
				METRIC_SOURCE {	 AEIP.MANAGED_ELEMENTS.STAGE_ME.countProblematicNodes }
				DESCRIPTION {"counts nodes with problems in the MARF stage"}	
				VALUE { 0 }
				THRESHOLD_CLASS {	Integer [0] } // valid only when holding 0 value
			}
		}
	}
}
\end{Verbatim}
\normalsize

\section{ASSL Code Specification for DMARF Self-Protection}
\label{appdx:assl-self-protection-spec}

What follows is the complete initial {\assl} specification of
the self-protection aspect for the {\dmarf}'s pipeline stages 
\cite{dmarf-assl-self-protection}.

\scriptsize
\begin{Verbatim}
// ASSL self-protecting specification model for DMARF

AS DMARF {

	TYPES {	ProxyCertificate }

	ASSELF_MANAGEMENT {
		// if a private message is detected as being insecure then
		// ignore it - the AE cannot neither receive nor resend it
		SELF_PROTECTING {
		  // a new incoming message has been detected
			FLUENT inSecurityCheck {
				INITIATED_BY { EVENTS.publicMessageIsComing }
				TERMINATED_BY {	EVENTS.publicMessageSecure,
						EVENTS.publicMessageInsecure }
			}
			MAPPING {
				CONDITIONS { inSecurityCheck }
				DO_ACTIONS { ACTIONS.checkPublicMessage }
			}
		}
	} // ASSELF_MANAGEMENT

	ACTIONS {
		ACTION checkPublicMessage {
			GUARDS { ASSELF_MANAGEMENT.SELF_PROTECTING.inSecurityCheck }
			VARS { Boolean senderIdentified }
			DOES {
				senderIdentified = false;
				FOREACH member in AES {
					IF ( NOT senderIdentified ) THEN
						senderIdentified =
							call member.ACTIONS.checkSenderCertificate
							(ASIP.MESSAGES.publicMessage.senderSignature)
					END
				};
				IF NOT senderIdentified THEN
// makes the metric invalid and thus, triggers the attached event
// and blocks all the operations with public messages
					set METRICS.thereIsInsecurePublicMessage.VALUE = true
				END
				ELSE
// makes the metric valid and thus, triggers the attached event
// and unblocks all the operations with public messages
					set METRICS.thereIsInsecurePublicMessage.VALUE = false
				END
			}
			ONERR_DOES {
				// if error then treat the message as insecure
				set METRICS.thereIsInsecurePublicMessage.VALUE = true
			}
		}
   } // ACTIONS

	EVENTS { // these events are used in the fluents specification
		EVENT publicMessageIsComing {
			ACTIVATION { SENT { ASIP.MESSAGES.publicMessage } }
		}
		EVENT publicMessageInsecure {
			GUARDS { NOT METRICS.thereIsInsecurePublicMessage }
			ACTIVATION { CHANGED { METRICS.thereIsInsecurePublicMessage } }
		}
		EVENT publicMessageSecure {
			GUARDS { METRICS.thereIsInsecurePublicMessage }
			ACTIVATION { CHANGED { METRICS.thereIsInsecurePublicMessage } }
		}

	} // EVENTS

	METRICS {
		// set to true when a new public insecure message
		// has been discovered
		METRIC thereIsInsecurePublicMessage {
			METRIC_TYPE { QUALITY	}
			DESCRIPTION {"detects an insecure message in the AE"}
			VALUE { false }
			// valid only when holding false value
			THRESHOLD_CLASS {	Boolean [false] }
		}
	}
} // AS DMARF

ASIP {
	MESSAGES {
		MESSAGE publicMessage {
			SENDER { ANY }
			RECEIVER { AES.STAGE_AE }
			MSG_TYPE {	 TEXT }
			PARAMETERS {	ProxyCertificate senderSignature }
		}
	}

	CHANNELS {
		CHANNEL publicLink {
			ACCEPTS {	MESSAGES.publicMessage }
			ACCESS { SEQUENTIAL }
			DIRECTION { INOUT }
		}
	}

	FUNCTIONS {
		//receive public messages if the message is secure
		FUNCTION receivePublicMessages {
			DOES {
				IF ( AS.METRICS.thereIsInsecurePublicMessage ) THEN
					MESSAGES.publicMessage << CHANNELS.publicLink
				END
			}
		}
		//send public messages if the message is secure
		FUNCTION sendPublicMessages {
			DOES {
				IF ( AS.METRICS.thereIsInsecurePublicMessage ) THEN
					MESSAGES.publicMessage >> CHANNELS.publicLink
				END
			}
		}
	}
}

AES {

	AE STAGE_AE {

		AESELF_MANAGEMENT {
			// if a private message is detected as being insecure then
			// ignore it - the AE cannot neither receive nor resend it
			SELF_PROTECTING {
				FLUENT inSecurityCheck {
					INITIATED_BY { EVENTS.privateMessageIsComming }
					TERMINATED_BY {	EVENTS.privateMessageSecure,
											EVENTS.privateMessageInsecure }
				}
				MAPPING {
					CONDITIONS { inSecurityCheck }
					DO_ACTIONS { ACTIONS.checkPrivateMessage }
				}
			}
		} // AESELF_MANAGEMENT

		AEIP {
			MESSAGES {
				MESSAGE privateMessage {
					SENDER { ANY }
					RECEIVER { AES.STAGE_AE }
					MSG_TYPE {	 TEXT }
					PARAMETERS {	ProxyCertificate senderSignature}
				}
			}

			CHANNELS {
				CHANNEL privateLink {
					ACCEPTS {	AEIP.MESSAGES.privateMessage }
					ACCESS { SEQUENTIAL }
					DIRECTION { INOUT }
				}
			}

			FUNCTIONS {
				//receive private messages if the message is secure
				FUNCTION receivePrivateMessages {
					DOES {
						IF ( AES.STAGE_AE.METRICS.thereIsInsecurePrivateMessage ) THEN
							AEIP.MESSAGES.privateMessage << AEIP.CHANNELS.privateLink
						END
					}
				}
				//send private messages if the message is secure
				FUNCTION sendPrivateMessages {
					DOES {
						IF ( AES.STAGE_AE.METRICS.thereIsInsecurePrivateMessage ) THEN
							AEIP.MESSAGES.privateMessage >> AEIP.CHANNELS.privateLink
						END
					}
				}
			}

			MANAGED_ELEMENTS {
				MANAGED_ELEMENT STAGE_ME {
					// checks if a node certificate is valid
					INTERFACE_FUNCTION checkNodeCertificate {
						PARAMETERS { ProxyCertificate	theCertificate }
						RETURNS { Boolean	}
					}
				}
			}
		} // AEIP

		ACTIONS {
			ACTION checkSenderCertificate {
				PARAMETERS {	ProxyCertificate theCertificate }
				RETURNS { Boolean }
				VARS { Boolean found }
				DOES {
					found = call AEIP.MANAGED_ELEMENTS.STAGE_ME.checkNodeCertificate
						(theCertificate);
					return found
				}
			}

			ACTION checkPrivateMessage {
				GUARDS { AESELF_MANAGEMENT.SELF_PROTECTING.inSecurityCheck }
				VARS { Boolean senderIdentified }
				DOES {
					senderIdentified = call ACTIONS.checkSenderCertificate
						( AEIP.MESSAGES.privateMessage.senderSignature );
					IF NOT senderIdentified THEN
						// makes the metric invalid and thus, triggers the attached event
						// and blocks all the operations with private messages
						set METRICS.thereIsInsecurePrivateMessage.VALUE = true
					END
					ELSE
						// makes the metric valid and thus, triggers the attached event
						// and unblocks all the operations with private messages
						set METRICS.thereIsInsecurePrivateMessage.VALUE = false
					END
				}
				ONERR_DOES {
					// if error then treat the message as insecure
					set METRICS.thereIsInsecurePrivateMessage.VALUE = true
				}
			}
		} // ACTIONS

		EVENTS {
			EVENT privateMessageIsComming {
				ACTIVATION { SENT { AEIP.MESSAGES.privateMessage } }
			}
			EVENT privateMessageInsecure {
				GUARDS { NOT METRICS.thereIsInsecurePrivateMessage }
				ACTIVATION { CHANGED { METRICS.thereIsInsecurePrivateMessage } }
			}
			EVENT privateMessageSecure {
				GUARDS { METRICS.thereIsInsecurePrivateMessage }
				ACTIVATION { CHANGED { METRICS.thereIsInsecurePrivateMessage } }
			}
		} // EVENTS

		METRICS {
			// set to true when an insecure private message is discovered
			METRIC thereIsInsecurePrivateMessage {
				METRIC_TYPE { QUALITY	}
				DESCRIPTION {"detects an insecure message in the AE"}
				VALUE { false }
				// valid only when holding false value
				THRESHOLD_CLASS {	Boolean [false] }
			}
		}
	}
}
\end{Verbatim}
\normalsize

\section{ASSL Code Specification for DMARF Self-Optimization}
\label{appdx:assl-self-optimization-spec}

What follows is the complete initial {\assl} specification of
the self-optimization aspect for the {\dmarf}'s pipeline stages 
\cite{dmarf-assl-self-optimization}.

\scriptsize
\begin{Verbatim}
// ASSL self-optimization specification model for DMARF 

AS DMARF {  

	ASSELF_MANAGEMENT {
		// DMARF strives to optimize by synchronizing cached 
		// results before starting with the Classification Stage 
		SELF_OPTIMIZING {
		  // DMARF enters in the Classification Stage 
			FLUENT inClassificationStage  {
				INITIATED_BY { EVENTS.enteringClassificationStage }
				TERMINATED_BY {	EVENTS.optimizationSucceeded, 
						EVENTS.optimizationNotSucceeded } 
			}
			MAPPING {
				CONDITIONS { inClassificationStage } 
				DO_ACTIONS { ACTIONS.runGlobalOptimization }  
			}
		}  
	} // ASSELF_MANAGEMENT
	
	ASARCHITECTURE {
		AELIST {AES.CLASSF_STAGE_NODE_1, AES.CLASSF_STAGE_NODE_2}
		DIRECT_DEPENDENCIES { 
			DEPENDENCY AES.CLASSF_STAGE_NODE_1 { AES.CLASSF_STAGE_NODE_2 }
			DEPENDENCY AES.CLASSF_STAGE_NODE_2 { AES.CLASSF_STAGE_NODE_1 }
		}
		GROUPS {
			GROUP CLASSF_STAGE { 
				MEMBERS { AES.CLASSF_STAGE_NODE_1, AES.CLASSF_STAGE_NODE_2 } 
			}
		}	
	}

	ACTIONS {
		ACTION runGlobalOptimization {	
			GUARDS { ASSELF_MANAGEMENT.SELF_OPTIMIZING.inClassificationStage }
			DOES {	
				FOREACH member IN ASARCHITECTURE.GROUPS.CLASSF_STAGE.MEMBERS {									
					call IMPL member.ACTIONS.synchronizeResults 
				}
			}
			TRIGGERS {					
				EVENTS.optimizationSucceeded	
			}
			ONERR_TRIGGERS {		
				// if error then report unsuccessful optimization 
				EVENTS.optimizationNotSucceeded		
			}
		} 
   } // ACTIONS

	EVENTS { // these events are used in the fluents specification 
		EVENT enteringClassificationStage { } 
		EVENT optimizationSucceeded { }
		EVENT optimizationNotSucceeded { }
	} // EVENTS

} // AS DMARF

AES {

	AE CLASSF_STAGE_NODE_1 {	
	
		AESELF_MANAGEMENT {
			SELF_OPTIMIZING {
				FLUENT inCPAdaptation {
					INITIATED_BY { AS.EVENTS.enteringClassificationStage }
					TERMINATED_BY {	EVENTS.cpAdaptationSucceeded, 
							EVENTS.cpAdaptationNotSucceeded } 
				}
				MAPPING {
					CONDITIONS { inCPAdaptation } 
					DO_ACTIONS { ACTIONS.adaptCP }  
				}
			}
		}

		ACTIONS {
			ACTION IMPL synchronizeResults {	
				GUARDS { AS.ASSELF_MANAGEMENT.SELF_OPTIMIZING.
						inClassificationStage 
				}
			}
			ACTION IMPL adaptCP {						
				GUARDS { AESELF_MANAGEMENT.SELF_OPTIMIZING.inCPAdaptation }
				TRIGGERS { EVENTS.cpAdaptationSucceeded }
				ONERR_TRIGGERS { EVENTS.cpAdaptationNotSucceeded }
			}
		} // ACTIONS

		EVENTS { // these events are used in the fluents specification 
			EVENT cpAdaptationSucceeded { }
			EVENT cpAdaptationNotSucceeded { }
		} // EVENTS
	}

	AE CLASSF_STAGE_NODE_2 {	
	
		AESELF_MANAGEMENT {
			SELF_OPTIMIZING {
				FLUENT inCPAdaptation {
					INITIATED_BY { AS.EVENTS.enteringClassificationStage }
					TERMINATED_BY {	EVENTS.cpAdaptationSucceeded, 
							EVENTS.cpAdaptationNotSucceeded } 
				}
				MAPPING {
					CONDITIONS { inCPAdaptation } 
					DO_ACTIONS { ACTIONS.adaptCP }  
				}
			}
		}

		ACTIONS {
			ACTION IMPL synchronizeResults {	
				GUARDS { AS.ASSELF_MANAGEMENT.SELF_OPTIMIZING.
						inClassificationStage 
				}
			}
			ACTION IMPL adaptCP {						
				GUARDS { AESELF_MANAGEMENT.SELF_OPTIMIZING.inCPAdaptation }
				TRIGGERS { EVENTS.cpAdaptationSucceeded }
				ONERR_TRIGGERS { EVENTS.cpAdaptationNotSucceeded }
			}
		} // ACTIONS

		EVENTS { // these events are used in the fluents specification 
			EVENT cpAdaptationSucceeded { }
			EVENT cpAdaptationNotSucceeded { }
		} // EVENTS}
	}
}
\end{Verbatim}
\normalsize


\end{document}